\documentclass[aps,showpacs,two column,showkeys,superscriptaddress,preprintnumbers]{revtex4-2}
\usepackage{epsfig,amssymb,subfigure,bm,dsfont}
\usepackage{chngcntr}
\usepackage{amsmath}
\usepackage{gensymb}
\usepackage{txfonts}
\usepackage[titletoc]{appendix}
\usepackage{amssymb}
\usepackage{array}
\usepackage{mathrsfs}
\usepackage{subfigure,overpic}
\usepackage{dcolumn}
\usepackage{graphicx}
\usepackage{epstopdf}
\usepackage{float}
\usepackage{balance}
\usepackage{lipsum}
\usepackage[colorlinks,
            linkcolor=blue,
            anchorcolor=blue,
            citecolor=blue,
            urlcolor=blue]{hyperref}
\usepackage{setspace}
\usepackage{multirow}
\usepackage{makecell}
\usepackage[normalem]{ulem} 
\usepackage{xcolor}
\usepackage{pifont}
\usepackage{graphicx}
\usepackage{tikz}     
\usepackage{chngcntr}
\usepackage{enumitem}
\usepackage[capitalize]{cleveref}
\usetikzlibrary{decorations.pathmorphing}
\definecolor{myblue}{rgb}{0,0,1}
\definecolor{myred}{rgb}{1,0,0}
\definecolor{myblack}{rgb}{0,0,0}
\definecolor{mymagenta}{rgb}{1,0,1}

\begin{document}

\title{RKKY interaction as a probe of valley-dependent spin splitting and odd-parity nature in Floquet collinear magnets}

\author{Hou-Jian Duan}
\email{dhjphd@mailbox.gxnu.edu.cn}
\affiliation{College of Physical Science and Technology, Guangxi Normal University, Guilin, Guangxi 541004, China}
\author{Yong-Jia Wu}
\affiliation{Guangdong Basic Research Center of Excellence for Structure and Fundamental Interactions of Matter, Guangdong Provincial Key Laboratory of Quantum Engineering and Quantum Materials, School of Physics, South China Normal University, Guangzhou 510006, China}
\affiliation{Guangdong-Hong Kong Joint Laboratory of Quantum Matter, Frontier Research Institute for Physics, South China Normal University, Guangzhou 510006, China}
\author{Xiaoliang Xiao}
\affiliation{Guangdong Basic Research Center of Excellence for Structure and Fundamental Interactions of Matter, Guangdong Provincial Key Laboratory of Quantum Engineering and Quantum Materials, School of Physics, South China Normal University, Guangzhou 510006, China}
\affiliation{Guangdong-Hong Kong Joint Laboratory of Quantum Matter, Frontier Research Institute for Physics, South China Normal University, Guangzhou 510006, China}
\author{Ming-Xun Deng}
\affiliation{Guangdong Basic Research Center of Excellence for Structure and Fundamental Interactions of Matter, Guangdong Provincial Key Laboratory of Quantum Engineering and Quantum Materials, School of Physics, South China Normal University, Guangzhou 510006, China}
\affiliation{Guangdong-Hong Kong Joint Laboratory of Quantum Matter, Frontier Research Institute for Physics, South China Normal University, Guangzhou 510006, China}
\author{Mou Yang}
\affiliation{Guangdong Basic Research Center of Excellence for Structure and Fundamental Interactions of Matter, Guangdong Provincial Key Laboratory of Quantum Engineering and Quantum Materials, School of Physics, South China Normal University, Guangzhou 510006, China}
\affiliation{Guangdong-Hong Kong Joint Laboratory of Quantum Matter, Frontier Research Institute for Physics, South China Normal University, Guangzhou 510006, China}
\author{Rui-Qiang Wang}
\email{wangruiqiang@m.scnu.edu.cn}
\affiliation{Guangdong Basic Research Center of Excellence for Structure and Fundamental Interactions of Matter, Guangdong Provincial Key Laboratory of Quantum Engineering and Quantum Materials, School of Physics, South China Normal University, Guangzhou 510006, China}
\affiliation{Guangdong-Hong Kong Joint Laboratory of Quantum Matter, Frontier Research Institute for Physics, South China Normal University, Guangzhou 510006, China}

\begin{abstract}
Odd-parity magnets were recently proposed to emerge in collinear antiferromagnets (AFMs) via Floquet engineering, with valley-dependent spin splitting underlying the odd-parity spin polarization. This proposal brings about two key challenges: detecting the spin splitting to verify the generation mechanism of these magnets, and identifying such polarization to confirm the odd-parity nature. Here, we show that the Ruderman-Kittel-Kasuya-Yosida (RKKY) interaction provides a unified magnetic probe for both tasks. Taking collinear $f$-wave magnets as a representative example, we find that the RKKY interaction yields distinct magnetic signals of the spin splitting---including a magnetism reversal in the Heisenberg/Ising terms and a sign alternation of the Dzyaloshinskii-Moriya (DM) term---that enable clear discrimination of collinear $f$-wave magnets from other related AFMs. Moreover, the DM term exhibits an $f$-wave shape with odd-parity symmetry, satisfying $J^{\alpha\beta}_{DM}(\mathbf{R}) = -J^{\alpha\beta}_{DM}(C_{2q}\mathbf{R})$ ($q=3$), which directly reflects the odd-parity spin polarization $S_z(\mathbf{k}) = -S_z(C_{2q}\mathbf{k})$ in momentum space. This behavior persists in $p$-wave magnets ($q=1$), demonstrating the generality of our approach. Our work establishes the RKKY interaction as a versatile probe for detecting band features of collinear odd-parity magnets, with predictions accessible to existing experimental techniques such as spin-polarized scanning tunneling spectroscopy.
\end{abstract}

\maketitle



\section{Introduction}
Spintronics aims to harness the spin of electrons, with the goal of realizing devices that are faster and more energy-efficient than conventional electronics \cite{Incorvia1}. Current devices are primarily based on ferromagnets (FMs) or antiferromagnets (AFMs)---the former in commercial use, the latter under research---but both have inherent limitations. Specifically, FMs are susceptible to magnetic field interference and suffer from slow read-write operations as well as high power consumption; although AFMs are stable and fast in response, their magnetic order is difficult to probe and manipulate \cite{Bai2}. To overcome these bottlenecks, researchers are actively seeking alternative materials. Recently, unconventional magnets proposed by the groups of \v{S}mejkal and Hellenes have attracted widespread interest \cite{mejkal1,mejkal2,hellenes}. The potential of these magnets for spintronics arises from their unique properties: they combine the zero net magnetization of AFMs with spin splitting reminiscent of FMs, but unlike FMs, this splitting is momentum-dependent and nonrelativistic. These characteristics endow such magnets with great promise for ultrafast and ultrahigh-density information storage, positioning them as strong candidates for next-generation spintronics \cite{Bai1,Fukaya1,Song1,Chakraborty1,Yamada1,fu2025_11,fu2026_11,fu2026_12,fu2026_13}. 

Unconventional magnets fall into two categories: even-parity ($d$, $g$, $i$-wave) magnets and odd-parity ($p$, $f$-wave) magnets \cite{mejkal1,mejkal2,hellenes,liu2026_11}. For the former, a wealth of candidate materials have been predicted in both two-dimensional (2D) and three-dimensional systems \cite{Krempasky2024,Lee2024,Ding2024,Zeng2024,Reimers2024,Yang2025,Karube2022,Bose2022,Feng2022,Wang2023,Bai2023,lei2024,Liu2024,Zeng2024_1,Jiang2025,Zhang2025,Li2025_1}. In contrast, material realizations of odd-parity magnets remain scarce, with known prototypes including CeNiAsO \cite{hellenes,Chakraborty1}, Gd$_3$Ru$_4$Al$_{12}$ \cite{Yamada1}, and NiI$_2$ \cite{Song2025_1}. Early studies suggested that odd-parity magnets could only be realized in noncollinear magnetic configurations \cite{hellenes,Song2025_1,Yu2025}. However, recent works have proposed that such magnets can instead be engineered in collinear AFMs via circularly polarized light (CPL) \cite{Liu2026_1,Huang2026,Zhu2026_1,Li2026_1,zhang2026_11}. The underlying mechanism is that off-resonant CPL introduces a valley-dependent staggered potential in the system Hamiltonian, which couples to the antiferromagnetic term and induces spin splitting. Notably, previous studies of spin splitting have mainly focused on altermagnets or noncollinear odd-parity magnets \cite{Zeng2024,Reimers2024,Yang2025,Ezawa2025}, where the spin-degenerate bands are split along the momentum direction. In contrast, the splitting here occurs along the energy direction and is valley-dependent, serving as the origin of odd-parity spin polarization in collinear magnetic configurations. To verify the generation mechanism of collinear odd-parity magnets, the first step is to detect this splitting: how can it be measured? Answering this question can also distinguish these magnets from equilibrium AFMs in the absence of light.

Beyond detecting the splitting itself, an even more fundamental question arises: how can we unambiguously verify whether a given system hosts genuine odd-parity magnetism? Since odd-parity spin polarization constitutes the defining hallmark of such magnets, its direct observation is essential to confirm their intrinsic character \cite{hellenes}. This task is particularly pressing for collinear odd-parity magnets, which have only recently been proposed. Even for the comparatively well-studied noncollinear counterparts, conventional approaches encounter severe limitations. For instance, experimental \cite{Yamada1,zhou2026} and theoretical \cite{Ezawa2025} transport studies have observed signatures consistent with a $p$-wave shape, yet these measurements cannot directly probe the odd-parity spin polarization itself. A comparable limitation applies to single-impurity probes in CeNiAsO-type noncollinear magnets, where Friedel oscillations---even for a magnetic impurity---remain isotropic, analogous to those in Rashba metals, and are therefore unable to uncover the underlying odd-parity nature \cite{Sukhachov2024}. These findings highlight the shortcomings of existing methods, and the challenge is no less acute for collinear odd-parity magnets, strongly motivating the search for alternative detection schemes capable of simultaneously probing the spin splitting and the odd-parity nature.

To tackle the two questions above, we turn to the Ruderman-Kittel-Kasuya-Yosida (RKKY) interaction, an indirect exchange coupling between magnetic impurities mediated by itinerant electrons. This interaction has been established as a powerful tool for probing the intrinsic properties of materials \cite{27_1,27,28_1,28,29,30,31,32,33,34,35,36,37,38,39,40,41,42,43,44,44_1,44_2,duan2026_1,duan2026_2,yarmohammadi2026_2}. Here, we demonstrate that it can be further harnessed as a unified probe for two key features of collinear odd-parity magnets. By analyzing the RKKY response of collinear $f$-wave magnets, we extract magnetic signals of valley-dependent spin splitting: a magnetism reversal in the Heisenberg/Ising terms and a sign alternation of the DM term. These signals enable reliable discrimination of collinear $f$-wave magnets from other related AFMs. Furthermore, the directional dependence of the RKKY interaction reveals that the DM term exhibits an $f$-wave shape with odd-parity symmetry, directly reflecting the odd-parity spin polarization in momentum space. We further extend this analysis to $p$-wave magnets, corroborating the generality of our approach.

\section{Models and method}
\label{model}
\subsection{Low-energy model for collinear $f$-wave magnets}
\begin{figure*}[tbh]
\centering \includegraphics[width=0.99\textwidth]{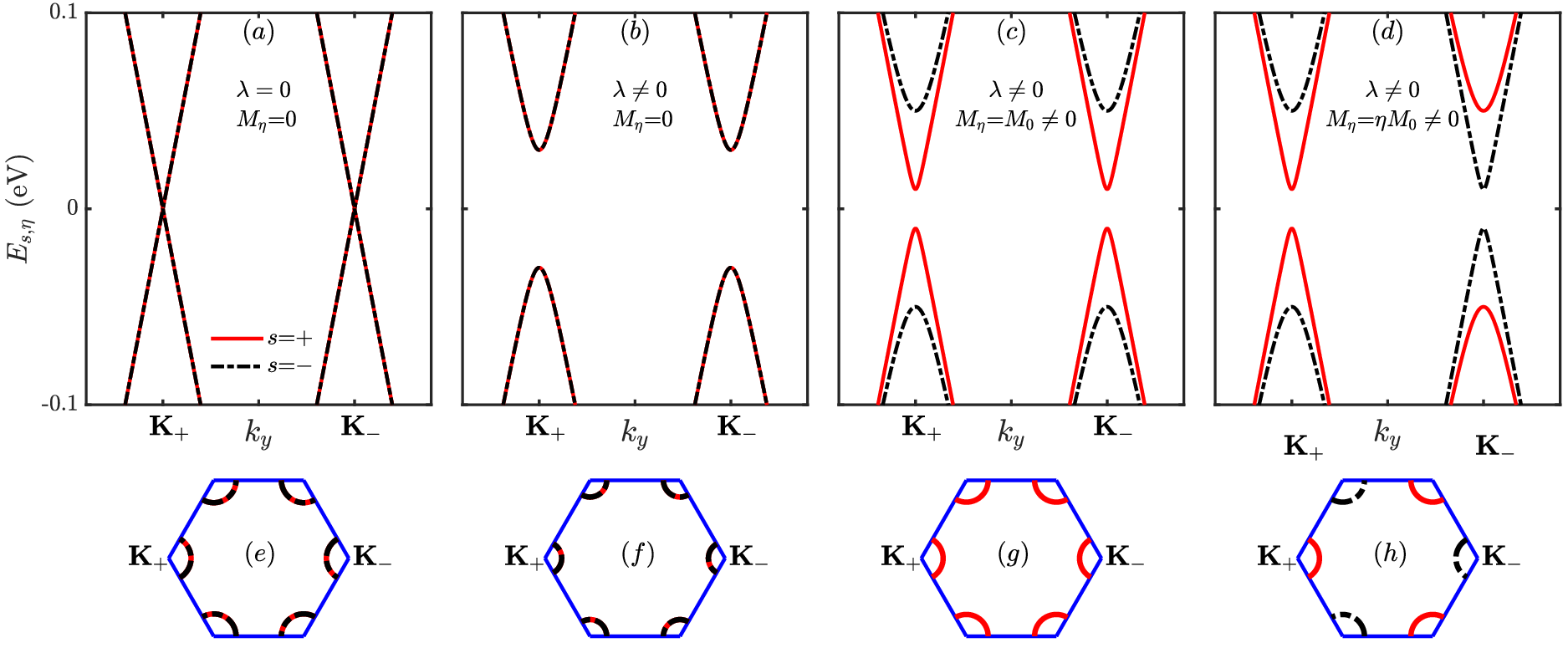}
\caption{$k_y$-axis dispersion (a$\sim$d) and Fermi surfaces at $u_F=0.045$ eV (e$\sim$h) for different systems: (a,e) graphene ($\lambda=0, M_\eta=0$), (b,f) collinear AFM ($\lambda\neq0, M_\eta=0$), (c,g) USS-AFM ($\lambda\neq0, M_\eta=M_0\neq0$), and (d,h) collinear $f$-wave magnet ($\lambda\neq0, M_\eta\neq0$). The nonzero value of $\lambda$ is taken as $\lambda=0.03t_0$, which is on the same order of magnitude as the parameters used in Ref.~\cite{Luo2019_1}, corresponding to the antiferromagnetic exchange energy in the graphene-based honeycomb lattice. The nonzero value of $M_0$ is set to $M_0=0.02t_0$, and we set $t_0=1$ eV for convenience.}
\label{fig1}
\end{figure*}
In this subsection, we briefly recall the realization of collinear $f$-wave magnets presented in Ref.~\cite{Zhu2026_1,Li2026_1}. Detailed derivations are given in Appendix \ref{model_pwave}; here we only summarize the key steps and results. We start from an unperturbed model of a two-dimensional collinear AFM with a honeycomb lattice. In the basis $\left( c_{\mathbf{k}A,\uparrow },c_{\mathbf{k}A,\downarrow },c_{\mathbf{k}B,\uparrow },c_{\mathbf{k}B,\downarrow }\right) ^{T}$, the tight-binding Hamiltonian of this model is given by
\begin{equation}\label{m1}
H_{0}\left( \mathbf{k}\right) =\left( 
\begin{array}{cc}
\lambda \sigma _{z} & t_0\sum_{i=1}^{3}e^{-i\mathbf{k}\cdot \mathbf{\delta }_{i}} \\ 
t_0\sum_{i=1}^{3}e^{i\mathbf{k}\cdot \mathbf{\delta }_{i}}& -\lambda \sigma _{z}%
\end{array}%
\right) ,
\end{equation}
where $a$ is the bond length, $\mathbf{\delta }_{1}=a\left( 1,0\right)$, $\mathbf{\delta }_{2,3}=a\left( -1/2,\pm \sqrt{3}/2\right)$, and $\sigma_{z}$ is the Pauli matrix acting on spin. In Eq.~(\ref{m1}), the off-diagonal terms describe nearest-neighbor hoppings, while the diagonal terms represent the antiferromagnetic term, which hosts magnetic order with the N\'eel vector along the $z$-axis. Such a collinear AFM can be realized in materials like $\rm MnPX_3$ ($\rm X = \rm S, Se$) \cite{Liu2020_1,Long2020}, graphene/antiferromagnetic-insulator heterostructures~\cite{Luo2019_1}, or functionalized tin films $ X$-$\rm{Sn}$ ($ X = \rm H, F, Cl, Br, I$) with a honeycomb lattice~\cite{Niu2017}. Because the Hamiltonian $H_{0}$ in Eq.~(\ref{m1}) preserves the combined inversion and time-reversal ($\mathcal{P}\mathcal{T}$) symmetry, the energy bands remain spin degenerate throughout the whole Brillouin zone.

To break the $\mathcal{P}\mathcal{T}$ symmetry, a beam of CPL is assumed to be normally incident on the 2D material. The corresponding vector potential is described as $\mathbf{A}=A_{0}\left[ \sin \left( \Omega t\right) ,\cos \left( \Omega t\right) \right]$ with period $T=2\pi/\Omega$. By applying the Peierls substitution $\mathbf{k\longrightarrow k+}e\mathbf{A/\hbar }$, the system Hamiltonian becomes time dependent, i.e., $H_{0}\left( \mathbf{k}\right)\longrightarrow H_{0}\left( \mathbf{k},t\right)$. Applying Floquet theory \cite{Fu2026_1,Fu2026_2,47,48,48_1,49,Cheng2025_1} and restricting to the high-frequency off-resonant regime, the driven system can be effectively captured by a static Hamiltonian that reads as
\begin{equation}\label{m2}
H(\mathbf{k})=V_{0}+\sum_{n\geq 1}\frac{\left[ V_{+n},V_{-n}\right] }{\hbar
\Omega}+O\left(\Omega ^{-2}\right),
\end{equation}
where $V_{n}=\frac{1}{T}%
\int_{0}^{T}H_0(\mathbf{k},t)e^{-in\hbar \Omega t}dt$. Keeping the leading-order contributions from $V_{0}$ and $V_{\pm 1}$ (the justification of this simplification is elaborated upon in Ref.~\cite{Liu2026_1}), the effective Hamiltonian $H(\mathbf{k})$ can be solved as
\begin{eqnarray}\label{m3}
H\left( \mathbf{k}\right) =\left( 
\begin{array}{cc}
t_0^{2}J_{1}^{2}\left( k_{a}a\right) f/\left(\hbar
\Omega\right) +\lambda \sigma _{z} & t_0J_{0}\left( k_{a}a\right) \sum_{i=1}^{3}e^{-i%
\mathbf{k}\cdot \mathbf{\delta }_{i}} \\ 
t_0J_{0}\left( k_{a}a\right) \sum_{i=1}^{3}e^{i\mathbf{k}\cdot \mathbf{\delta }%
_{i}} & -t_0^{2}J_{1}^{2}\left( k_{a}a\right) f /\left(\hbar \Omega\right) -\lambda \sigma _{z}%
\end{array}%
\right),
\end{eqnarray}
with $k_a = eA_0/\hbar$ and $f$ given by 
\begin{equation*} 
f =4\sqrt{3}\left[ \cos \left( \frac{3k_{x}a}{2}\right) -\cos \left( \frac{3k_{y}a}{2}\right) \right] \sin \left( \frac{%
\sqrt{3}k_{y}a}{2}\right),
\end{equation*}
where $J_{n}(x)$ denotes the $n$th-order Bessel function of the first kind.

Expanding the Hamiltonian of Eq. (\ref{m3}) around the valleys $\mathbf{K}_{\eta=\pm}=\left[0,-\eta 4\pi/(3\sqrt{3}a)\right]$ and considering the weak driving field ($k_a a \ll 1$), one obtains the effective low-energy Floquet Hamiltonian $H_{s,\eta}$ as
\begin{eqnarray}\label{m4}
H_{s,\eta }=v_{F}\left( \eta q_{y}\tau _{x}+q_{x}\tau
_{y}\right) +s\lambda \tau _{z}-M_\eta\tau _{z},
\end{eqnarray}
with 
\begin{eqnarray}\label{m5}
M_\eta=\eta M_0,
\end{eqnarray}
where $v_F=3at_0/2$, $M_{0}=\left( v_{F}k_{a}\right) ^{2}/\left( \hbar \Omega \right)$, $s=\pm$ for up and down spins, and $\tau_i$ are the Pauli matrices acting on the sublattice degree of freedom. The Hamiltonian in Eq.~(\ref{m4}) coincides exactly with that reported in Ref.~\cite{Zhu2026_1,Li2026_1}. The same type of valley-dependent $f$-wave spin splitting also emerges in compensated collinear magnets via sublattice currents \cite{Lin2025}.

The last term in Eq.~(\ref{m4}), which originates from the optical field, acts as a valley-dependent staggered sublattice potential. To demonstrate how this potential, together with the antiferromagnetic term, affects the band structure, we present the band diagrams for four systems (collinear $f$-wave magnet and other three baseline systems) in Fig.~\ref{fig1}. For $\lambda = 0$ and $M_\eta = 0$, the system reduces to that of pristine graphene, and the bands are spin-degenerate at all momenta [Fig.~\ref{fig1}(a)]. When the antiferromagnetic term is included ($\lambda \neq 0,M_\eta = 0$), a gap opens but the spin degeneracy remains intact [Fig.~\ref{fig1}(b)], as a consequence of the preserved $\mathcal{PT}$ symmetry. However, for the valley-dependent staggered potential  ($M_\eta = \eta M_0 \neq 0$), we find that it breaks the $\mathcal{PT}$ symmetry and consequently couples with the antiferromagnetic term to lift the spin degeneracy [Fig.~\ref{fig1}(d)]. Specifically, unlike the momentum-direction spin splitting in altermagnets or noncollinear odd-parity magnets, the spin-degenerate bands here are split along the energy direction in a valley-dependent manner. The system thus becomes an $f$-wave magnet, whose odd-parity spin polarization is characterized by the $C_{2q}$ rotational symmetry \cite{Fu2026_2}, satisfying $S_z(\mathbf{k}) = -S_z(C_{2q}\mathbf{k})$ ($q=3$), as seen in Fig.~\ref{fig1}(h). To illustrate the importance of this valley dependence, we now consider an alternative system with 
\begin{eqnarray}\label{m6}
M_\eta= M_0,
\end{eqnarray}
which corresponds to a valley-independent staggered potential \cite{Giovannetti2007,Zhou2007,Qiao2011}. Unlike the case with $M_\eta= \eta M_0 \neq 0$, the potential here couples with the antiferromagnetic term to induce spin splitting without valley dependence [Fig.~\ref{fig1}(c)]. In other words, the Fermi surfaces at different valleys carry the same spin [Fig.~\ref{fig1}(g)], and thus the system no longer exhibits odd parity. For simplicity, we refer to this system as the uniform-spin-splitting AFM (USS-AFM). In Sec.~\ref{III}-A, we will use RKKY interactions to characterize the valley-dependent spin splitting, and the extracted magnetic signals are expected to help verify the generation mechanism of collinear $f$-wave magnets and distinguish them from the other three baseline systems.

\subsection{RKKY interaction formalism}
The RKKY interaction is modeled by placing two magnetic impurities on the 2D collinear magnets, at positions $\mathbf{r}_1$ and $\mathbf{r}_2$. These impurities interact with the itinerant electrons of the host material via a contact potential $H_{i}=J_c\mathbf{S}_{i}\cdot \bm{\sigma}\,\delta(\mathbf{r}-\mathbf{r}_i)$, where $\mathbf{S}_i$ ($i=1,2$) denotes the spin of impurity. The scattering of itinerant electrons between the two impurities induces an effective indirect exchange interaction between them, known as the RKKY interaction. Following the standard perturbation theory \cite{51,52,53,54} and keeping $J_c$ to second order, this interaction can be expressed as
\begin{equation}\label{m7}
H_{R}^{\alpha\beta}=-\frac{J_{c}^{2}}{\pi }\mathrm{Im}\int_{-\infty }^{u_{F}}d\omega \mathrm{Tr}\left[ (\mathbf{S}_{1}\cdot \sigma )G^{\alpha\beta}(\mathbf{%
R},\omega)(\mathbf{S}_{2}\cdot \sigma )G^{\beta\alpha}(-\mathbf{R},\omega
)\right] ,
\end{equation}%
where $\mathbf{R} = \mathbf{r}_i - \mathbf{r}_j$, $u_F$ is the Fermi energy. In the above equation, $G^{\alpha\beta}$ is the real-space retarded Green's function of the clean system, which takes the form of a $2\times 2$ matrix in spin space. The subscripts $\alpha,\beta \in {A, B}$ indicate whether an impurity is located on the sublattice A or the sublattice B. Since the Hamiltonian $H_{s,\eta}$ of Eq. (\ref{m4}) is diagonal in spin space, the retarded Green's function $G^{\alpha\beta}(\pm\mathbf{R}, \omega)$ can also be expressed as a diagonal form
\begin{equation}\label{m8}
G^{\alpha\beta}(\pm\mathbf{R}, \omega)=\left(
                                           \begin{array}{cc}
                                             G^{\alpha\beta}_+(\pm\mathbf{R}, \omega) & 0 \\
                                             0 & G^{\alpha\beta}_-(\pm\mathbf{R}, \omega) \\
                                           \end{array}
                                         \right)
.
\end{equation}
Here the subscript $s=\pm$ in $G^{\alpha\beta}_s$ labels spin-up and spin-down, respectively.
\par
The matrix elements $G^{\alpha\beta}_s$ of the Green's function in Eq.~(\ref{m8}) can be obtained via the following formula,
\begin{eqnarray}\label{m9}
\begin{split}
&\left( 
\begin{array}{cc}
G_{s}^{AA}\left( \pm \mathbf{R},\omega \right)  & G_{s}^{AB}\left( \pm 
\mathbf{R},\omega \right)  \\ 
G_{s}^{BA}\left( \pm \mathbf{R},\omega \right)  & G_{s}^{BB}\left( \pm 
\mathbf{R},\omega \right) 
\end{array}%
\right) \\
&=\frac{1}{\left( 2\pi \right) ^{2}}\sum_{\eta }\int d\mathbf{q}\frac{1}{%
\omega +i0^{+}-H_{s,\eta }}e^{\pm i\left( \mathbf{q+K}_{\eta }\right) 
\mathbf{R}}.
\end{split}
\end{eqnarray}
After integrating over the momentum $\mathbf{q}=(q_x,q_y)$, we obtain the analytical expression for $G^{\alpha\beta}_s(\pm\mathbf{R},\omega)$ as
\begin{eqnarray}\label{m10}
\begin{split}
G^{AA}_s\left( \pm \mathbf{R},\omega \right) &=\frac{-1}{2\pi v_{F}^{2}}%
\sum_{\eta }e^{\pm i\mathbf{K}_{\eta }\mathbf{R}}\left( \omega +s\lambda
-M_\eta\right) g_{s,\eta}, \\
G^{BB}_s\left( \pm \mathbf{R},\omega \right)& =\frac{-1}{2\pi v_{F}^{2}}%
\sum_{\eta }e^{\pm i\mathbf{K}_{\eta }\mathbf{R}}\left( \omega -s\lambda
+M_\eta\right) g_{s,\eta}, \\
G^{AB}_s\left( \pm \mathbf{R},\omega \right) &=\frac{\mp 1}{2\pi v_{F}^{2}}%
\sum_{\eta }e^{\pm i\mathbf{K}_{\eta }\mathbf{R}}e^{i\eta \theta _{R}}h_{s,\eta}, \\
G^{BA}_s\left( \pm \mathbf{R},\omega \right) &=\frac{\pm 1}{2\pi v_{F}^{2}}%
\sum_{\eta }e^{\pm i\mathbf{K}_{\eta }\mathbf{R}}e^{-i\eta \theta _{R}}h_{s,\eta},
\end{split}
\end{eqnarray}
with 
\begin{eqnarray}\label{m11}
\begin{split}
g_{s,\eta}&=K_{0}\left\{ \frac{R}{v_{F}\sqrt{1/\left[ \left( s\lambda -M_\eta\right)
^{2}-\omega ^{2}\right] }}\right\} ,\\
h_{s,\eta}&=\frac{K_{1}\left\{ \frac{R}{v_{F}\sqrt{1/\left[ \left( s\lambda
-M_\eta\right) ^{2}-\omega ^{2}\right] }}\right\} }{\sqrt{1/\left[ \left(
s\lambda -M_\eta\right) ^{2}-\omega ^{2}\right] }},
\end{split}
\end{eqnarray}
where $K_n(x)$ ($n = 0, 1$) denotes the $n$th-order modified Bessel function of the second kind.

Inserting the Green's function $G^{\alpha\beta}(\pm\mathbf{R}, \epsilon)$ from Eq.~(\ref{m8}) into Eq.~(\ref{m7}) and tracing over the spin degrees of freedom, the RKKY interaction can be expressed in the following form
\begin{eqnarray}\label{m12}
\begin{split}
H_{R}^{\alpha \beta }=J_{H}^{\alpha \beta }\mathbf{S}_{1}\cdot \mathbf{S}_{2}+J_{I}^{\alpha
\beta }S_{1}^{z}S_{2}^{z}+J_{DM}^{\alpha \beta }\left( \mathbf{S}_{1}\times 
\mathbf{S}_{2}\right) _{z},
\end{split}
\end{eqnarray}
with
\begin{eqnarray}\label{m13}
\begin{split}
J_{H}^{\alpha \beta }=\frac{-J_c^{2}}{\pi }{\rm{Im}}\int\nolimits_{-\infty }^{0}\left[ \sum\limits_{j=\pm }G_{j}^{\alpha \beta
}\left( \mathbf{R},\omega \right) G_{-j}^{\beta \alpha }\left( -\mathbf{R}%
,\omega \right) \right] d\omega  , \\
J_{I}^{\alpha \beta }=\frac{-J_c^{2}}{\pi }{\rm{Im}}\int\nolimits_{-\infty
}^{0}\left[ \sum\limits_{j=\pm }G_{j}^{\alpha \beta }\left( \mathbf{R}%
,\omega \right) G_{j}^{\beta \alpha }\left( -\mathbf{R},\omega \right) %
\right] d\omega -J_{H}^{\alpha \beta }, \\
J_{DM}^{\alpha \beta }=\frac{J_c^{2}}{\pi }{\rm{Im}}\int\nolimits_{-\infty
}^{0}i\left[ \sum\limits_{j=\pm }jG_{j}^{\alpha \beta }\left( \mathbf{R}%
,\omega \right) G_{-j}^{\beta \alpha }\left( -\mathbf{R},\omega \right) %
\right] d\omega .
\end{split}
\end{eqnarray}
In Eq.~(\ref{m12}), $J_{H}^{\alpha\beta}$ denotes the Heisenberg term and $J_{I}^{\alpha\beta}$ is the Ising term. Both terms couple collinear spins, but they differ in that the former describes the isotropic part of the RKKY interaction whereas the latter accounts for the anisotropic part. The last term $J_{DM}^{\alpha\beta}$ represents the DM interaction, which is considered to be the origin of the anomalous Hall effect on the surface of topological insulators \cite{27_1}. It is worth noting that both the Ising term and the DM term are aligned with the $z$-axis, a feature that originates from the choice of the N\'eel vector along the $z$-direction. In this respect, the Ising term behaves analogously to that in $d$-wave altermagnets \cite{42,43,44,duan2026_1}.

\begin{figure*}[tbh]
\centering \includegraphics[width=0.98\textwidth]{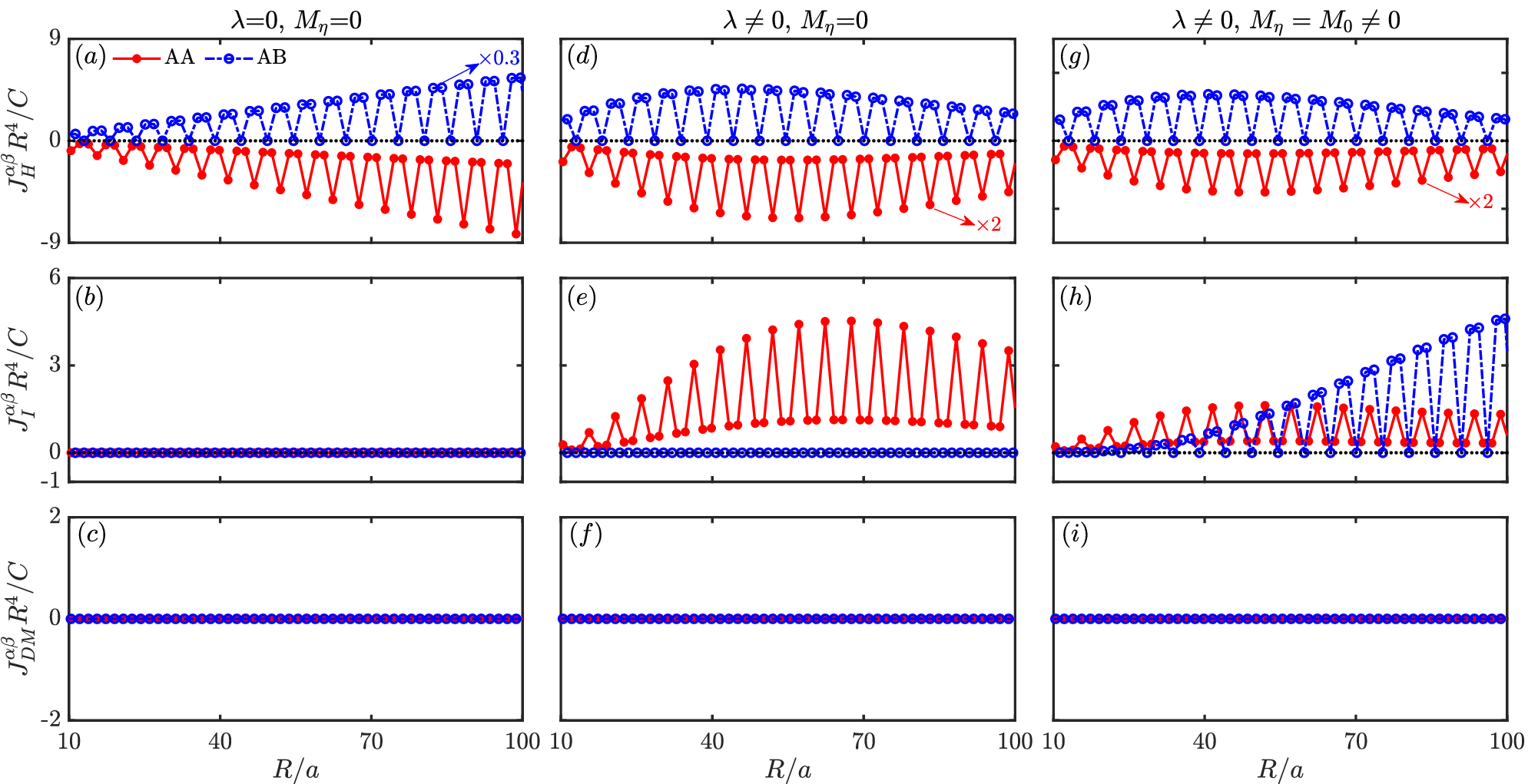}
\caption{RKKY components $J_i^{\alpha\beta}$ ($i=H,I,DM$) as a function of the impurity distance $R$ for three baseline systems: (a$\sim$c) graphene ($\lambda = 0, M_\eta = 0$), (d$\sim$f) the collinear AFM ($\lambda \neq 0, M_\eta = 0$), and (g$\sim$i) the USS-AFM ($\lambda \neq 0, M_\eta = M_0 \neq 0$). Solid and open circles represent the RKKY interactions for impurities placed on the same and different sublattices, respectively. Here, $C=J_c^2/(2\pi)$ and $u_F=0$.}
\label{fig2}
\end{figure*}

\section{Results and discussion}
\label{III}
This section addresses two distinct aspects of the RKKY response in the collinear $f$-wave magnet. The first part is dedicated to extracting RKKY signals of valley-dependent spin splitting. The second part turns to the characterization of the odd-parity spin polarization. Together, these two analyses establish the RKKY interaction as a comprehensive probe of the band structure of collinear odd-parity magnets. Throughout all discussions that follow, we restrict ourselves to the zero-Fermi-energy case ($u_F = 0$), so as to prevent the oscillations induced by a finite Fermi energy from disturbing the magnetic signals of interest. In addition, we mainly focus on the long-range impurity configurations (i.e., relatively large $Ra$), where the impurities are assumed to have negligible influence on the low-energy band structure, as discussed in Ref.~\cite{Shiranzaei2018}.

\subsection{RKKY signals distinguishing collinear $f$-wave magnet from other systems}
We proceed in two steps. We first examine the RKKY interaction in three baseline systems (graphene, the collinear AFM, and the USS-AFM). Then, we study the RKKY interaction in the collinear $f$-wave magnet and compare it with the three baseline systems. Here, the impurities are placed along the zigzag direction (i.e., the $y$-axis), which is well suited for probing the band properties of the collinear $f$-wave magnet.

\subsubsection{RKKY interaction in graphene, collinear AFM, and USS-AFM}
Substituting Eqs.~(\ref{m10}) and~(\ref{m11}) into Eq.~(\ref{m13}), we calculate the RKKY interaction for three baseline systems, as displayed in Fig.~\ref{fig2}. We now discuss them in turn.
\par
\begin{enumerate}[label=(\arabic*)]
 \item For $\lambda = 0$ and $M_\eta = 0$ (pristine graphene), only the Heisenberg term $J^{\alpha\beta}_H$ survives, as shown in Figs.~\ref{fig2}(a$\sim$c). As the impurity distance $R$ varies, both $J^{AA}_H$ and $J^{AB}_H$ oscillate. The key distinction is that $J^{AA}_H$ is always negative (ferromagnetic), while $J^{AB}_H$ is positive (antiferromagnetic). These findings are fully consistent with earlier reports on graphene \cite{Saremi2007,Black-Schaffer2010,Sherafati2011,Kogan2011}.
\par
\item For the collinear AFM ($\lambda \neq 0$ and $M_\eta = 0$), two notable differences from the graphene case can be identified in Figs.~\ref{fig2}(d$\sim$f). First, the interaction decays more rapidly with increasing $R$, a consequence of the gap opened by the finite $\lambda$ [Fig.~\ref{fig1}(b)]. A similar gap-induced rapid decay has also been observed in phosphorene \cite{Duan2017}. Second, an Ising term $J^{AA}_I$ emerges when the impurities occupy the same sublattice, driven by the antiferromagnetic term ($\lambda \tau_z \sigma_z$). As seen from Eq.~(\ref{m10}), once $\lambda \neq 0$, $G^{AA}$ develops a component proportional to $q_z\sigma_z$, which couples with $q_0\sigma_0$ to give rise to an Ising term oriented along the $z$-axis. In contrast, $G^{AB} \propto \sigma_0$ remains independent of $\lambda$, so no Ising term appears when the impurities are on different sublattices.
\par
 \item Turning to the USS-AFM ($\lambda \neq 0$ and $M_\eta = M_0 \neq 0$), the most significant distinction from the previous two systems is that a nonzero Ising term $J^{AB}_I$ arises when the impurities are on different sublattices. This behavior originates from the spin splitting of the USS-AFM [Fig.~\ref{fig1}(d)], which leads to $h_+ \neq h_-$ [see Eq.~(\ref{m10})] and thus introduces a $\sigma_z$-dependent contribution to $G^{AB}$, in direct analogy to the appearance of $J^{AA}_I$ in the collinear AFM.
\end{enumerate}
\par
Beyond the differences discussed above, the RKKY interaction in the three baseline systems also has two common aspects. First, neither the antiferromagnetic term ($\lambda\neq0$), nor the valley-independent staggered potential ($M_\eta = M_0 \neq 0$), nor their combination can generate a DM term. This is guaranteed by the $\mathcal{P}$ symmetry of the Green's function, i.e., $G^{\alpha\beta}(\mathbf{R}, \omega) = G^{\beta\alpha}(-\mathbf{R}, \omega)$. Second, the magnetism of $J^{\alpha\beta}_H$ and $J^{\alpha\beta}_I$ remains unchanged regardless of the impurity distance $R$. Taking $J^{\alpha\beta}_H$ as an example, as $R$ varies, $J^{AA}_H$ is always ferromagnetic (negative), while $J^{AB}_H$ is always antiferromagnetic (positive), as shown in Figs.~\ref{fig2}(a, d, g).
\par
To facilitate comparison with the collinear $f$-wave magnet, we now analyze how intravalley and intervalley contributions determine the single-type magnetism of $J^{\alpha\beta}_H$ and $J^{\alpha\beta}_I$ in the three baseline systems. Taking $J^{AA}_H$ as an example, substitution of Eq.~(\ref{m10}) into Eq.~(\ref{m13}) allows $J^{AA}_H$ to be written as
\begin{eqnarray}\label{m14}
\begin{split}
J_{H}^{AA }&=J_{H}^{AA,{\rm intra} }+J_{H}^{AA,{\rm inter}}\cos\left[\left(\mathbf{K}_+-\mathbf{K}_-\right) \mathbf{R}\right],\\
&=J_{H}^{AA,{\rm intra} }J_{\rm sign} , \\
\end{split}
\end{eqnarray}
with 
\begin{eqnarray}\label{m15}
\begin{split}
J_{\rm sign}=1+\left(J_{H}^{AA,{\rm inter} }/J_{H}^{AA,{\rm intra} }\right)\cos\left[\left(\mathbf{K}_+-\mathbf{K}_-\right)\mathbf{R}\right].
\end{split}
\end{eqnarray}
In the first line of Eq.~(\ref{m14}), the first term is the non-oscillatory intravalley contribution, while the second term describes the intervalley contribution, which consists of an amplitude $J_{H}^{AA,{\rm inter}}$ multiplied by the oscillatory factor $\cos[(\mathbf{K}_+-\mathbf{K}_-) \mathbf{R}]$. The intravalley contribution $J_{H}^{AA,{\rm intra}}$ is always negative due to particle-hole symmetry, as previously discussed for graphene \cite{Sherafati2011}. Consequently, the sign of $J_{H}^{AA}$ is controlled entirely by $J_{\rm sign}$ [Eq.~(\ref{m15})]. For the three baseline systems, the weights of the intravalley and intervalley contributions are strictly equal ($J_{H}^{AA,{\rm inter}}/J_{H}^{AA,{\rm intra}}=1$), so $J_{\rm sign}=1+\cos[(\mathbf{K}_+-\mathbf{K}_-)\mathbf{R}]$, which repeats the sequence $2, 1/2, 1/2$ as $R$ varies. This ensures that $J_{H}^{AA}$ always retains the sign of $J_{H}^{AA,{\rm intra}}$ and thus remains purely ferromagnetic. As we will show in the next subsection, this behavior changes qualitatively when the valley-dependent spin splitting modifies the ratio $J_{H}^{AA,{\rm inter}}/J_{H}^{AA,{\rm intra}}$, leading to a profound impact on the magnetism of the collinear RKKY components.

\subsubsection{RKKY signals as a probe of valley-dependent spin splitting in collinear $f$-wave magnets}
\begin{figure}[tbh]
	\centering \includegraphics[width=0.47\textwidth]{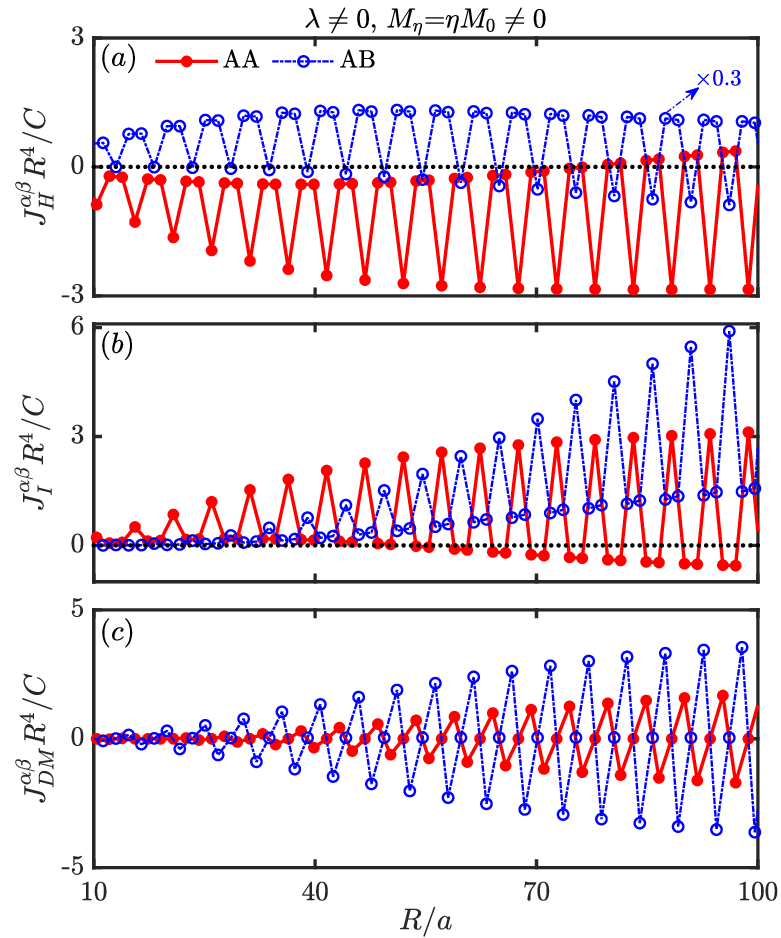}
	\caption{RKKY components as a function of impurity distance in the collinear $f$-wave magnet ($\lambda \neq 0, M_\eta = \eta M_0 \neq 0$). Solid and open circles represent the RKKY interactions for impurities placed on the same and different sublattices, respectively.}
	\label{fig3}
\end{figure}

\begin{figure}[tbh]
\centering \includegraphics[width=0.47\textwidth]{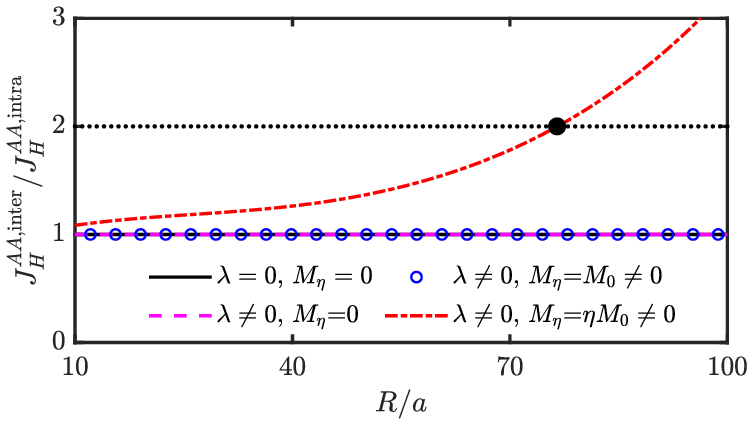}
\caption{Spatial dependence of the ratio $J^{AA,{\rm inter}}_H / J^{AA,{\rm intra}}_H$ for different systems. The parameters for each system are indicated in the legend.}
\label{fig4}
\end{figure}

\begin{figure}[tbh]
\centering \includegraphics[width=0.49\textwidth]{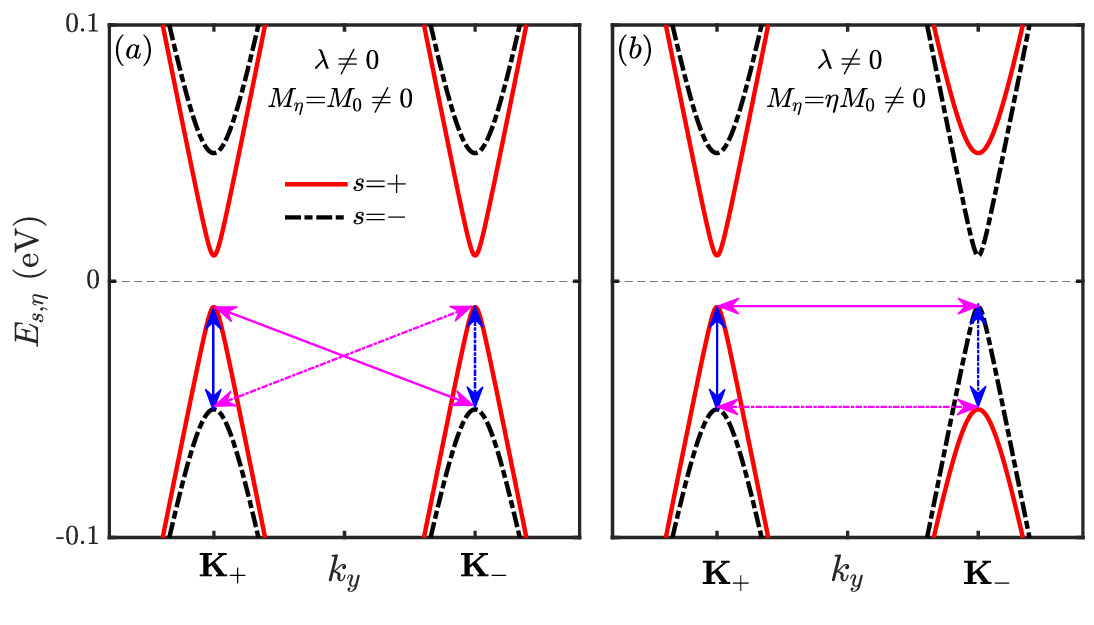}
\caption{$k_y$-axis dispersion for (a) the USS-AFM ($\lambda\neq0, M_\eta=M_0\neq0$) and (b) the collinear $f$-wave magnet ($\lambda\neq0, M_\eta=\eta M_0\neq0$). The blue arrows (solid and dashed) schematically represent the intravalley contribution, while the magenta ones (solid and dashed) schematically denote the intervalley contribution.}
\label{fig5}
\end{figure}

The valley-dependent spin splitting is a key band feature in the generation mechanism of collinear $f$-wave magnets (Fig.~\ref{fig1}), and it is this feature that gives rise to a unique RKKY response. This response, in turn, can be used to probe the splitting itself. We now show how the distinct RKKY signals arising from this splitting---particularly the reversal of magnetism and the emergence of the DM term---serve to confirm this mechanism and distinguish the collinear $f$-wave magnet from other related systems.

\paragraph{Reversal of magnetism}
In the collinear $f$-wave magnet, the Heisenberg term $J^{\alpha\beta}_H$ and the Ising term $J^{AA}_I$ behave qualitatively differently from the baseline systems: they no longer exhibit a single type of magnetism (i.e., neither purely ferromagnetic nor purely antiferromagnetic), as shown in Figs.~\ref{fig3}(a,b). Focusing on $J^{AA}_H$ as a representative example (solid line in Fig.~\ref{fig3}(a)), we find that for small $R$ it remains ferromagnetic, similar to the baseline systems (Fig.~\ref{fig2}). Once $R$ exceeds a certain critical value, however, $J^{AA}_H$ undergoes a sign change at specific values of $R$, flipping from ferromagnetic to antiferromagnetic---in stark contrast to the behavior in Fig.~\ref{fig2}. Thus, this reversal, arising directly from the valley-dependent spin splitting, characterizes the splitting itself and distinguishes the collinear $f$-wave magnet from other AFMs (the collinear AFM and USS-AFM).
\par
A central quantity underlying this reversal is the ratio of the intervalley and intravalley contributions (i.e., $J^{AA,{\rm inter}}_H / J^{AA,{\rm intra}}_H$), whose variation across different systems is shown in Fig.~\ref{fig4}. In the three baseline systems---where spin splitting is either absent or valley-independent---this ratio is always exactly 1, guaranteeing the single-type (ferromagnetic) magnetism of $J^{AA}_H$, as discussed previously. In the collinear $f$-wave magnet, by contrast, the ratio $J^{AA,{\rm inter}}_H / J^{AA,{\rm intra}}_H$ grows rapidly with increasing $R$ (see Fig.~\ref{fig4}). To understand how this affects the magnetism of $J^{AA}_H$, we recall that the sign of $J^{AA}_H$ in Eq.~(\ref{m14}) is governed by $J_{\rm sign} = 1 + \left(J_{H}^{AA,{\rm inter}} / J_{H}^{AA,{\rm intra}}\right) \cos[(\mathbf{K}_+ - \mathbf{K}_-) \mathbf{R}]$, with the oscillatory factor repeating the sequence $1, -1/2, -1/2$ as $R$ varies. Consequently, when the ratio $J^{AA,{\rm inter}}_H / J^{AA,{\rm intra}}_H$ grows continuously from 1 and surpasses 2, the sequence of $J_{\rm sign}$ evolves from $2, 1/2, 1/2$ to $>3, <0, <0$. This means that the magnetism of $J^{AA}_H$ reverses at two out of every three values of $R$, a process that is assisted by the oscillatory factor.
\par
To understand the mechanism behind this growing ratio, we substitute Eq.~(\ref{m10}) into Eq.~(\ref{m13}) and find that $J^{AA}_H$ consists of four terms: two terms form the intravalley contribution (schematically indicated by the blue arrows in Fig.~\ref{fig5}), and the other two form the intervalley contribution (magenta arrows). Physically, each arrow connects two valleys, and the magnitude of each term is determined by how close those valleys are to the Fermi energy ($u_F=0$): the closer the valleys, the larger the term. Applying this rule to the two systems:
\begin{enumerate}[label=(\arabic*)]
 \item In the USS-AFM [Fig.~\ref{fig5}(a)], the spin splitting is valley-independent, so the valleys connected by the blue arrows and those connected by the magenta arrows lie equally distant from the Fermi energy. By the rule above, equal distances imply equal contributions---which is precisely why $J^{AA,{\rm inter}}_H = J^{AA,{\rm intra}}_H$ for all $R$.
 \item The situation is different in the collinear $f$-wave magnet [Fig.~\ref{fig5}(b)], where the spin splitting is valley-dependent: the spin polarization takes opposite signs at $\mathbf{K}_+$ and $\mathbf{K}_-$, so that the valleys connected by the solid magenta arrow (intervalley) are pushed closer to the Fermi energy than those connected by the blue arrows (intravalley). As a result, the intervalley contribution becomes dominant ($J^{AA,{\rm inter}}_H / J^{AA,{\rm intra}}_H>1$). Moreover, since larger $R$ enhances the dominance of electrons near the Fermi energy in the RKKY interaction, the intervalley contribution---originating from valleys closer to the Fermi energy---grows stronger with $R$, while the intravalley contribution weakens. Consequently, the ratio $J^{AA,{\rm inter}}_H / J^{AA,{\rm intra}}_H$ increases with $R$.
\end{enumerate}
\par
\begin{figure*}[tbh]
\centering \includegraphics[width=0.99\textwidth]{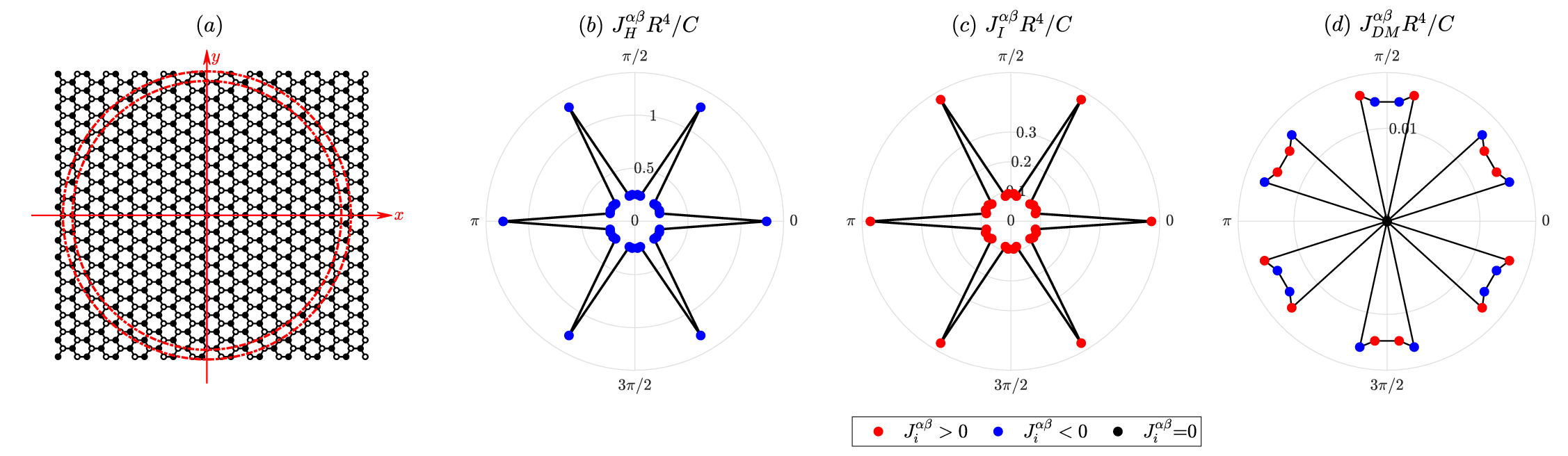}
\caption{(a) Impurity configuration: one impurity is fixed at the origin, while the other is allowed to move among the lattice sites enclosed by the red ring. (b)--(d) RKKY components $J_i^{\alpha\beta}$ ($i=H,I,DM$) obtained for the configuration shown in (a). In (b)--(d), the angle of the solid circle relative to the $x$-axis represents the azimuthal angle of the impurities, while the distance from the origin represents the magnitude of the RKKY component; blue (red) corresponds to $J^{\alpha\beta}_i<0$ ($J^{\alpha\beta}_i>0$), and black denotes $J^{\alpha\beta}_i=0$.}
\label{fig6}
\end{figure*}
\paragraph{Emergence of the DM term}
In the collinear $f$-wave magnet, the DM term $J^{\alpha\beta}_{DM}$ is nonzero and alternates in sign with increasing $R$ (Fig.~\ref{fig3}(c)), while it vanishes in the three baseline systems. This term thus provides another magnetic signal that characterizes the valley-dependent spin splitting and distinguishes the $f$-wave magnet from these systems. Notably, the DM term here is purely $z$-component, in marked contrast to two-dimensional Rashba SOC systems, where only in-plane DM terms exist \cite{Imamura2004_1,Mross2009}. It also differs from the $z$-component DM term in even-parity magnets, which arises from the cooperation between altermagnetism and Rashba SOC \cite{duan2026_1}. To understand the physical origin of this unique DM term, we analyze its emergence from two complementary perspectives:
\begin{enumerate}[label=(\arabic*)]
 \item From Eq.~(\ref{m13}), we find that a nonzero $J^{\alpha\beta}_{\rm DM}$ requires $G_j \neq G_{-j}$---i.e., spin splitting must be present. Beyond this prerequisite, once the splitting acquires valley dependence, the $\mathcal{P}$ symmetry of the Green's function is broken, leading to $G_s^{\alpha\beta}(\mathbf{R}, \omega) \neq G_s^{\beta\alpha}(-\mathbf{R}, \omega)$; this asymmetry is known to generate the DM term \cite{32}. To see this explicitly, we consider $G_s^{AA}(\pm\mathbf{R}, \omega)$ from Eq.~(\ref{m10}), which can be rewritten as
\begin{eqnarray*}
\begin{split}
G^{AA}_s\left( \pm \mathbf{R},\omega \right) \propto
\sum_{\eta }e^{ i\mathbf{K}_{\eta }\mathbf{R}}\left( \omega +s\lambda
-M_{\pm\eta}\right) g_{s,{\pm\eta}},
\end{split}
\end{eqnarray*}
where the valley dependence encoded in $M_\eta = \eta M_0$ leads to $\left( \omega +s\lambda-M_{\eta}\right) g_{s,{\eta}} \neq \left( \omega +s\lambda
-M_{-\eta}\right) g_{s,{-\eta}}$, giving rise to $G_s^{AA}(\mathbf{R}, \omega) \neq G_s^{AA}(-\mathbf{R}, \omega)$.
\item The band structure provides a more intuitive picture for the generation of the DM term. To verify this picture explicitly, we substitute Eq.~(\ref{m10}) into Eq.~(\ref{m13}) and find that $J^{AA}_{DM}$ consists solely of an intervalley contribution comprising two terms ($J_{DM}^{AA,\mathrm{inter1}}$ and $J_{DM}^{AA,\mathrm{inter2}}$), schematically represented by the solid and dashed magenta arrows in Fig.~\ref{fig5}, respectively. Their difference yields the DM term: $J_{DM}^{AA}=J_{DM}^{AA,\mathrm{inter1}}-J_{DM}^{AA,\mathrm{inter2}}$. In the USS-AFM, the valley independence of the spin splitting ensures that the valleys connected by the solid magenta arrow and those connected by the dashed magenta arrow lie equally distant from the Fermi energy. Following the earlier rule that equal distances imply equal contributions, the two terms cancel each other out, and no DM term is generated. In the collinear $f$-wave magnet, by contrast, the valley-dependent spin splitting pushes the valleys connected by the solid magenta arrow (corresponding to $J_{DM}^{AA,\mathrm{inter1}}$) closer to the Fermi energy, while those connected by the dashed arrow ($J_{DM}^{AA,\mathrm{inter2}}$) are shifted farther away. Thus, the two terms no longer cancel, and the DM interaction emerges.
\end{enumerate}

\subsection{RKKY signals characterizing the odd-parity spin polarization}
A defining feature of odd-parity magnets is odd-parity spin polarization in momentum space, i.e., $S_z(\mathbf{k}) = -S_z(C_{2q}\mathbf{k})$, where $q=1$ for $p$-wave magnets and $q=3$ for $f$-wave magnets, and $S_z$ denotes the spin polarization along the $z$-axis. Having established that the RKKY response can detect the valley-dependent spin splitting, we now turn to the second question: can the same interaction also reveal the odd-parity spin polarization? We will show that the DM term can serve as a probe of this unique spin polarization, regardless of whether the system is a $f$- or $p$-wave magnet.
\par
To capture this spin polarization, we study the azimuthal-angle dependence of the RKKY interaction in real space for the collinear $f$-wave magnet. Here, we consider the impurity configuration shown in Fig.~\ref{fig6}(a), where one impurity is fixed at the origin and the other is allowed  to move among the lattice sites enclosed by the red ring. The corresponding RKKY interaction is displayed in Figs.~\ref{fig6}(b$\sim$d), where the distance of each solid circle from the origin gives the magnitude of $J^{\alpha\beta}_i$; red (blue) and black indicate $J^{\alpha\beta}_i > 0$ ($<0$) and $J_i=0$, respectively. From these figures, we see a striking difference between the Heisenberg/Ising terms ($J^{\alpha\beta}_H$ or $J^{\alpha\beta}_I$) and the DM term ($J^{\alpha\beta}_{DM}$): although both exhibit an $f$-wave shape, only the DM term shows odd parity, i.e., $J^{\alpha\beta}_{DM}(\mathbf{R}) = -J^{\alpha\beta}_{DM}(C_{2q}\mathbf{R})$ ($q=3$). This distinction originates from the oscillatory factors: the Heisenberg/Ising terms involve a cosine factor $\cos[(\mathbf{K}_+ - \mathbf{K}_-)\mathbf{R}]$, while the DM term features a sine factor $\sin[(\mathbf{K}_+ - \mathbf{K}_-)\mathbf{R}]$, which changes sign under the $C_{2q}$ rotation of $\mathbf{R}$. This directly reflects the odd-parity spin polarization in momentum space. We further verify that the impurity positions need not be restricted to the ring in Fig.~\ref{fig6}(a); other impurity configurations merely enrich the spatial pattern of the DM term, without compromising its ability to probe the odd-parity spin polarization. 
\par
Finally, we extend our study to CeNiAsO-type $p$-wave magnets and find analogous results: the DM term not only captures the $p$-wave shape but also probes the odd-parity spin polarization (see Appendix~\ref{RKKY_pwave}), demonstrating the generality of our approach. This stands in stark contrast to conventional transport measurements, which can reveal the $p$-wave shape but fail to resolve the odd-parity nature \cite{Ezawa2025,zhou2026,Sukhachov2024}. In addition, Friedel oscillations induced by a single impurity have also been investigated in such systems, yet they remain isotropic and thus cannot probe this polarization either \cite{Sukhachov2024}. By contrast, the RKKY interaction studied here arises from scattering between two impurities. This two-terminal nature enables it to sense the anisotropy of the spin-split bands, much like radar detection (which loses its effect for a single impurity), thereby providing a viable route to probing the odd-parity spin polarization.

\section{Summary}
\label{summary}
In summary, we have investigated the RKKY interaction in collinear $f$-wave magnets as a representative example of Floquet-engineered odd-parity magnets. Our central finding is that the RKKY interaction serves as a unified magnetic probe capable of addressing two distinct but equally essential questions: detecting the valley-dependent spin splitting to verify the generation mechanism of such magnets, and characterizing the odd-parity spin polarization to confirm the odd-parity nature. Specifically, the RKKY response yields rich magnetic signals of valley-dependent spin splitting---a magnetism reversal in the Heisenberg/Ising terms and a sign alternation of the DM term---that enable clear discrimination from other related AFMs. Moreover, the DM term exhibits an $f$-wave shape with odd-parity symmetry, satisfying $J^{\alpha\beta}_{DM}(\mathbf{R}) =-J^{\alpha\beta}_{DM}(-C_{2q}\mathbf{R})$ ($q=3$), which directly reflects the odd-parity spin polarization in momentum space. This behavior extends to $p$-wave magnets ($q=1$), demonstrating the generality of our approach and overcoming the limitations of single-impurity probes and conventional transport measurements. Our predictions are accessible to existing experimental techniques such as spin-polarized scanning tunneling spectroscopy \cite{59,60}, capable of detecting magnetization curves of individual atoms, or electron spin resonance combined with optical detection \cite{61}. Overall, this work establishes the RKKY interaction as a versatile and experimentally feasible probe for detecting the band features of collinear odd-parity magnets.

\acknowledgements
This work was supported by the National Natural Science Foundation of China (Grants No. 12574050, No. 12274146, No. 11904107, No. 11774100, No. 12504073), by the Guangdong Basic and Applied Basic Research Foundation under Grant No. 2023B1515020050, by the Guangdong NSF of China (Grant No. 2026A1515012155).

\section*{Data Availability} 
The data that support the findings of this article are not publicly available. The data are available from the authors upon reasonable request.

\newpage
\begin{widetext}
\appendix
\renewcommand\thesection{\Roman{section}}
\renewcommand\thesubsection{\Alph{subsection}}
\def\CTeXPreproc{Created by ctex v0.2.12, don't edit!}
\numberwithin{equation}{section}
\counterwithin{figure}{section}  
\renewcommand{\thefigure}{S\arabic{figure}}
\setcounter{figure}{0}

\section{collinear $f$-wave magnets induced by off-resonant CPL in collinear AFMs with a honeycomb lattice}
\label{model_pwave}
We start from an unperturbed model of a two-dimensional collinear AFM with a honeycomb lattice \cite{Zhu2026_1,Li2026_1}. In the basis $\left( c_{\mathbf{k}A,\uparrow },c_{\mathbf{k}A,\downarrow },c_{\mathbf{k}B,\uparrow },c_{\mathbf{k}B,\downarrow }\right) ^{T}$, the tight-binding Hamiltonian of this model is given by
\begin{equation}\label{I1}
H_{0}\left( \mathbf{k}\right) =\left( 
\begin{array}{cc}
0 & t_0\sum_{i=1}^{3}e^{-i\mathbf{k}\cdot \mathbf{\delta }_{i}} \\ 
t_0\sum_{i=1}^{3}e^{i\mathbf{k}\cdot \mathbf{\delta }_{i}}& 0
\end{array}\right) \sigma_0+\left( \begin{array}{cc}
\lambda \sigma _{z} & 0 \\ 
0& -\lambda \sigma _{z}%
\end{array}%
\right) ,
\end{equation}
where $\mathbf{\delta }_{1}=a\left( 1,0\right)$, $\mathbf{\delta }_{2,3}=a\left( -1/2,\pm \sqrt{3}/2\right) $, and $\sigma_{z}$ is the Pauli matrix acting on spin. In Eq. (\ref{I1}), the off-diagonal terms describe nearest-neighbor hoppings, while the diagonal terms represent the antiferromagnetic term. Here, a beam of CPL is assumed to be injected in the $z$ axis. The corresponding vector potential is described as $\mathbf{A}(t)=A_0[\sin(\Omega t),\cos(\Omega t)]$ with period $T=2\pi/\Omega$.  By applying the Peierls substitution $\mathbf{k}\rightarrow \mathbf{k}+e\mathbf{A}/\hbar$, the system Hamiltonian becomes time-dependent. Using the Floquet theory \cite{Fu2026_1,Fu2026_2,47,48,48_1,49,Cheng2025_1} with the off-resonant condition of $\hbar\Omega\gg BW$ ($BW$ is the bandwidth), the modified part of the Hamiltonian induced by
light reads as
\begin{equation}
\label{I2}
H=V_{0}+\sum_{n\geq 1}\frac{\left[ V_{+n},V_{-n}\right] }{n\hbar
\Omega}+O\left(\Omega ^{-2}\right),
\end{equation}
where $V_{n}=\frac{1}{T}%
\int_{0}^{T}H_0(\mathbf{k}+e\mathbf{A}/\hbar)e^{-in\hbar \Omega t}dt$. Specifically, $V_0$ can be calculated as
\begin{eqnarray}
\label{I3}
\begin{split}
V_0=&\frac{1}{T}\int_{0}^{T}H_{0}\left( \mathbf{k}+e\mathbf{A}/\hbar
\right) dt, \\
=&\frac{1}{T}\int_{0}^{T}\left( 
\begin{array}{cc}
\lambda \sigma _{z} & 0 \\ 
0 & -\lambda \sigma _{z}%
\end{array}%
\right) dt+\frac{1}{T}\int_{0}^{T}t_0\left( 
\begin{array}{cc}
0 & e^{-i\left[ k_{x}+eA_{0}\sin \left( \Omega t\right) /\hbar \right] a} \\ 
e^{i\left[ k_{x}+eA_{0}\sin \left( \Omega t\right) /\hbar \right] a} & 0%
\end{array}%
\right) \sigma_0 dt, \\
&+\frac{1}{T}\int_{0}^{T}t_0\left( 
\begin{array}{cc}
0 & e^{i\left[ k_{x}+eA_{0}\sin \left( \Omega t\right) /\hbar \right] a/2-i%
\sqrt{3}\left[ k_{y}+eA_{0}\cos \left( \Omega t\right) /\hbar \right] a/2}
\\ 
e^{-i\left[ k_{x}+eA_{0}\sin \left( \Omega t\right) /\hbar \right] a/2+i%
\sqrt{3}\left[ k_{y}+eA_{0}\cos \left( \Omega t\right) /\hbar \right] a/2} & 
0%
\end{array}%
\right)\sigma_0 dt,\\
&+\frac{1}{T}\int_{0}^{T}t_0\left( 
\begin{array}{cc}
0 & e^{i\left[ k_{x}+eA_{0}\sin \left( \Omega t\right) /\hbar \right] a/2+i%
\sqrt{3}\left[ k_{y}+eA_{0}\cos \left( \Omega t\right) /\hbar \right] a/2}
\\ 
e^{-i\left[ k_{x}+eA_{0}\sin \left( \Omega t\right) /\hbar \right] a/2-i%
\sqrt{3}\left[ k_{y}+eA_{0}\cos \left( \Omega t\right) /\hbar \right] a/2} & 
0%
\end{array}%
\right)\sigma_0 dt.
\end{split}
\end{eqnarray}
Since $\mathbf{A}(t)$ is a periodic function of time $t$, the integrand in the above equation can be simplified as
\begin{eqnarray}
\label{I4}
\begin{split}
&\frac{1}{T}\int_{0}^{T}e^{\pm ik_a a\left[ \sin \left(
\Omega t\right) \pm \sqrt{3}\cos \left( \Omega t\right) \right] /2}dt\\
&=\frac{1}{T}\int_{0}^{T}e^{\pm ik_a a\sin \left( \Omega
t\pm \pi /3\right) }dt ,\\
&=\frac{1}{T}\int_{0}^{T}e^{\pm ik_a a\sin \left( \Omega
t\right) }dt,\\
&=J_{0}\left( k_a a\right),
\end{split}
 \end{eqnarray}
where $k_a=eA_{0}a/\hbar$, $J_n(x)$ denotes the $n$th-order Bessel function of the first kind. Substituting the result of the above integration into Eq.~(\ref{I3}), $V_0$ can be solved as
\begin{eqnarray}
\label{I5}
\begin{split}
V_{0}=\left( 
\begin{array}{cc}
0 & J_{0}\left( k_a a\right) t_0\sum_{i=1}^{3}e^{-i\mathbf{k}%
\cdot \mathbf{\delta }_{i}} \\ 
J_{0}\left( k_a a\right) t_0\sum_{i=1}^{3}e^{i\mathbf{k}\cdot 
\mathbf{\delta }_{i}} & 0%
\end{array}%
\right) \sigma _{0}+\left( 
\begin{array}{cc}
\lambda \sigma _{z} & 0 \\ 
0 & -\lambda \sigma _{z}%
\end{array}%
\right).
\end{split}
 \end{eqnarray}
The above results show that, in contrast to the Hamiltonian $H_0$ in Eq.~(\ref{I1}), the light-induced correction enters $V_0$ solely via the replacement $t_0 \rightarrow t_0 J_{0}\left( k_a a\right)$. Similarly, one can obtain other Floquet sidebands as
\begin{eqnarray}
\label{I6}
\begin{split}
V_{n}=tJ_{n}\left( k_{a}a\right) \left[ \left( 
\begin{array}{cc}
0 & e^{-i\mathbf{k}\cdot \mathbf{\delta }_{1}}\left( -1\right) ^{n} \\ 
e^{i\mathbf{k}\cdot \mathbf{\delta }_{1}} & 0%
\end{array}%
\right) +e^{-in\pi /3}\left( 
\begin{array}{cc}
0 & e^{-i\mathbf{k}\cdot \mathbf{\delta }_{2}} \\ 
e^{i\mathbf{k}\cdot \mathbf{\delta }_{2}}\left( -1\right) ^{n} & 0%
\end{array}%
\right) +e^{in\pi /3}\left( 
\begin{array}{cc}
0 & e^{-i\mathbf{k}\cdot \mathbf{\delta }_{3}} \\ 
e^{i\mathbf{k}\cdot \mathbf{\delta }_{3}}\left( -1\right) ^{n} & 0%
\end{array}%
\right) \right] \sigma _{0}.
\end{split}
\end{eqnarray}
\par
Under the high-frequency off-resonant condition, we only need to keep the leading-order contributions from $V_{0}$ and $V_{\pm 1}$ (as justified in Ref.~\cite{Liu2026_1}). Substituting $V_{0}$ and $V_{\pm 1}$ into Eq.~(\ref{I2}) yields
\begin{eqnarray}\label{I7}
H\left( \mathbf{k}\right) =\left( 
\begin{array}{cc}
t_0^{2}J_{1}^{2}\left( k_{a}a\right)f/\left(\hbar
\Omega\right)  & t_0J_{0}\left( k_{a}a\right) \sum_{i=1}^{3}e^{-i%
\mathbf{k}\cdot \mathbf{\delta }_{i}} \\ 
t_0J_{0}\left( k_{a}a\right) \sum_{i=1}^{3}e^{i\mathbf{k}\cdot \mathbf{\delta }%
_{i}} & -t_0^{2}J_{1}^{2}\left( k_{a}a\right) f /\left(\hbar \Omega\right) 
\end{array}%
\right)\sigma_0+\left(
          \begin{array}{cc}
            \lambda \sigma _{z} & 0 \\
            0 & -\lambda \sigma _{z} \\
          \end{array}
        \right)
,
\end{eqnarray}
with 
\begin{eqnarray*}
f=4\sqrt{3}\left[ \cos \left( \frac{3k_{x}a}{2}\right) -\cos \left( \frac{3k_{y}a}{2}\right) \right] \sin \left( \frac{\sqrt{3}k_{y}a}{2}\right).
\end{eqnarray*}
Considering the weak driving field ($k_a a \ll 1$), the Bessel function $J_0(x)$ and $J_1(x)$ can be approximated as
\begin{eqnarray}\label{I8}
J_0(x)\approx 1, \;\; J_1(x)\approx x/2.
\end{eqnarray}
Under the above approximation, the Hamiltonian $H(\mathbf{k})$ in Eq.~(\ref{I7}) can be further simplified as
\begin{eqnarray}\label{I9}
H\left( \mathbf{k}\right) =\left( 
\begin{array}{cc}
t_0^{2}k_{a}^2a^2f/\left(4\hbar
\Omega\right)  & t_0 \sum_{i=1}^{3}e^{-i%
\mathbf{k}\cdot \mathbf{\delta }_{i}} \\ 
t_0\sum_{i=1}^{3}e^{i\mathbf{k}\cdot \mathbf{\delta }%
_{i}} & -t_0^{2}k_a^2a^2 f /\left(4\hbar \Omega\right) 
\end{array}%
\right)\sigma_0+\left(
          \begin{array}{cc}
            \lambda \sigma _{z} & 0 \\
            0 & -\lambda \sigma _{z} \\
          \end{array}
        \right).
\end{eqnarray}
Expanding the above Hamiltonian around the valleys $\mathbf{K}_{\eta=\pm}=\left[0,-\eta 4\pi/(3\sqrt{3}a)\right]$, one obtains the effective low-energy Floquet Hamiltonian $H_{s,\eta}$ as
\begin{eqnarray}\label{I10}
H_{s,\eta }=v_{F}\left( \eta q_{y}\tau _{x}+q_{x}\tau
_{y}\right) +s\lambda \tau _{z}-M_\eta\tau _{z},
\end{eqnarray}
where $v_F=3at_0/2$, $M_\eta=\eta M_0$ with $M_{0}=\left( v_{F}k_{a}\right) ^{2}/\left( \hbar \Omega \right)$, $s=\pm$ for up and down spins, and $\tau_i$ is the Pauli matrix in the sublattice space. The above Hamiltonian coincides exactly with that reported in Ref.~\cite{Zhu2026_1,Li2026_1}. The last term in $H_{s,\eta }$, which originates from the optical field, acts as a valley-dependent staggered sublattice potential.
\par
Here, the driving frequency $\hbar\Omega $ and the bandwidth $BW$ are set as $\hbar\Omega=1{\rm eV} $ and $BW=0.1{\rm eV}$. The off-resonant condition is satisfied since $\hbar\Omega \gg BW$. The setting of the bandwidth, as well as the frequency, is reasonable since we only concern the behavior of electrons near zero Fermi energy ($u_F = 0$). In addition, for $M_0 = 0.2t_0$ (the value used for nonzero $M_0$ in the main text), we have $k_a a \approx 0.067$, which satisfies the weak-driving-field assumption (i.e., $k_a a \ll 1$).

\section{ RKKY interactions in the CeNiAsO-type $p$-wave magnets}
\label{RKKY_pwave}

Here, we start from the low-energy Hamiltonian of the CeNiAsO-type $p$-wave magnet~\cite{Chakraborty1,Ezawa2025}:
\begin{eqnarray}\label{I11}
\begin{split}
H^\prime(\mathbf{k})&=Dk^{2}\sigma _{0}+J_0k_{y}\sigma _{z}, \\
&=\left( 
\begin{array}{cc}
Dk_{x}^{2}-\frac{J_0^{2}}{4D}+D\left( k_{y}+\frac{J_0}{2D}\right) ^{2} & 0 \\ 
0 & Dk_{x}^{2}-\frac{J_0^{2}}{4D}+D\left( k_{y}-\frac{J_0}{2D}\right) ^{2}%
\end{array}%
\right). 
\end{split}
\end{eqnarray}
As can be seen from the second line of the above equation, two parabolic bands with opposite spins are split along the $k_y$-axis.

\begin{figure*}[tbh]
\centering \includegraphics[width=0.9\textwidth]{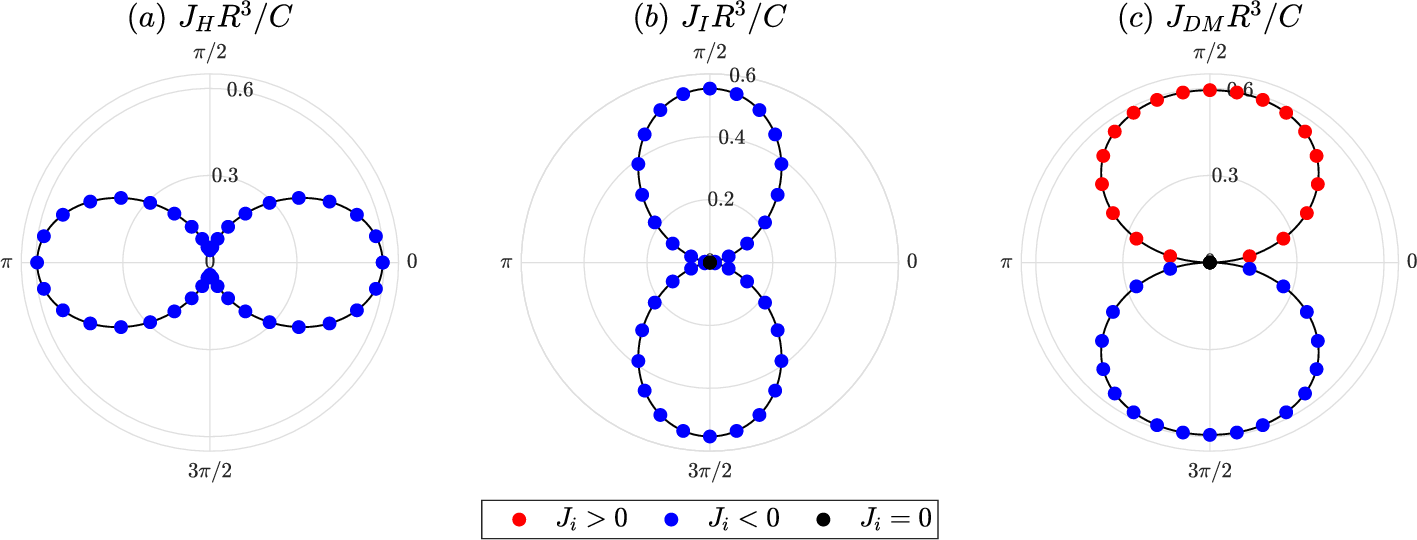}
\caption{RKKY components (a) $J_H$, (b) $J_I$, and (c) $J_{DM}$ as functions of the azimuthal angle of the impurities with $u_F=0$. The distance of each solid circle from the origin represents the magnitude of the RKKY component; blue (red) corresponds to $J_i<0$ ($J_i>0$), and black denotes $J_i=0$.}
\label{figS2}
\end{figure*}
\par
Substituting the Hamiltonian $H^\prime(\mathbf{k})$ from Eq.~(\ref{I11}) into the following retarded Green's function
\begin{eqnarray}\label{I12}
\begin{split}
 G\left( \pm \mathbf{R,}\omega \right) =\frac{1}{\left( 2\pi \right)
^{2}}\int d\mathbf{k}\frac{1}{\omega +i0^{+}-H^{\prime }\left( \mathbf{k}%
\right) }e^{\pm i\left( k_{x}R_{x}+k_{y}R_{y}\right) },
\end{split}
\end{eqnarray}
we can rewrite the real-space Green's function $G(\pm \mathbf{R},\omega)$ as
\begin{eqnarray}\label{I13}
\begin{split}
G\left( \pm \mathbf{R,}\omega \right) &=\frac{\sigma _{0}+\sigma
_{z}}{2\left( 2\pi \right) ^{2}}\int dk_{x}dk_{y}\frac{1}{\omega +i0^{+}-%
\left[ Dk_{x}^{2}-\frac{J_0^{2}}{4D}+D\left( k_{y}+\frac{J_0}{2D}\right) ^{2}%
\right] }e^{\pm i\left( k_{x}R_{x}+k_{y}R_{y}\right) }, \\
&+\frac{\sigma _{0}-\sigma _{z}}{2\left( 2\pi \right) ^{2}}\int dk_{x}dk_{y}%
\frac{1}{\omega +i0^{+}-\left[ Dk_{x}^{2}-\frac{J_0^{2}}{4D}+D\left( k_{y}-%
\frac{J_0}{2D}\right) ^{2}\right] }e^{\pm i\left( k_{x}R_{x}+k_{y}R_{y}\right).
}
\end{split}
\end{eqnarray}
In the first (second) line of the above equation, by making the substitution $k_{y}=k_{y}^{\prime }-J_0/2D$ ($k_{y}=k_{y}^{\prime }+J_0/2D$), we can further simplify $G(\pm \mathbf{R},\omega)$ to
\begin{eqnarray}\label{I14}
\begin{split}
G\left( \pm \mathbf{R,}\omega \right) =\frac{\sigma _{0}+\sigma
_{z}}{2\left( 2\pi \right) ^{2}}\int dk_{x}dk_{y}^{\prime }\frac{1}{\omega
^{+}-\left( Dk_{x}^{2}-\frac{J_0^{2}}{4D}+Dk_{y}^{\prime 2}\right) }e^{\pm
i\left( k_{x}R_{x}+k_{y}^{\prime }R_{y}-\frac{J_0R_{y}}{2D}\right) },\\ 
+\frac{\sigma _{0}-\sigma _{z}}{2\left( 2\pi \right) ^{2}}\int
dk_{x}dk_{y}^{\prime }\frac{1}{\omega ^{+}-\left( Dk_{x}^{2}-\frac{J_0^{2}}{4D}%
+Dk_{y}^{\prime 2}\right) }e^{\pm i\left( k_{x}R_{x}+k_{y}^{\prime }R_{y}+%
\frac{J_0R_{y}}{2D}\right) }.
\end{split}
\end{eqnarray}
Converting the Cartesian coordinates to polar coordinates, the above Green's function can be rewritten as
\begin{eqnarray}\label{I15}
\begin{split}
G\left( \pm \mathbf{R,}\omega \right) =\frac{\sigma _{0}+\sigma _{z}}{%
2\left( 2\pi \right) ^{2}}e^{\mp i\frac{J_0R_{y}}{2D}}\int kdkd\theta \frac{1}{%
\omega ^{+}-\left( Dk^{2}-\frac{J_0^{2}}{4D}\right) }e^{\pm ikR\cos \left(
\theta -\theta _{R}\right) },\\ 
+\frac{\sigma _{0}-\sigma _{z}}{2\left( 2\pi \right) ^{2}}e^{\pm i\frac{%
J_0R_{y}}{2D}}\int kdkd\theta \frac{1}{\omega ^{+}-\left( Dk^{2}-\frac{J_0^{2}}{%
4D}\right) }e^{\pm ikR\cos \left( \theta -\theta _{R}\right) }.
\end{split}
\end{eqnarray}
After successively integrating over $\theta$ and $k$ in the above equation, we obtain the analytical expression for $G(\pm \mathbf{R},\omega)$:
\begin{eqnarray}\label{I16}
\begin{split}
G\left( \pm \mathbf{R,}\omega \right) =-\frac{1}{4\pi D}K_{0}\left( \frac{R%
\sqrt{-J_0^{2}-4D\omega }}{2D}\right) \left[ e^{\mp \frac{iJ_0R_{y}}{2D}}\left(
\sigma _{0}+\sigma _{z}\right) +e^{\pm i\frac{J_0R_{y}}{2D}}\left( \sigma
_{0}-\sigma _{z}\right) \right].
\end{split}
\end{eqnarray}
\par
Substituting the above Green's function into the second-order perturbation formula in Eq.~(\ref{m7}) of the main text (with the superscripts $\alpha\beta$ dropped) and integrating over the energy $\omega$, we obtain the RKKY interaction in the following form:
\begin{eqnarray}\label{I17}
\begin{split}
H_{R}=J_{H}\mathbf{S}_{1}\cdot \mathbf{S}_{2}+J_{I}S_{1}^{z}S_{2}^{z}+J_{DM}\left( \mathbf{S}_{1}\times \mathbf{S}%
_{2}\right) _{z},
\end{split}
\end{eqnarray}
with
\begin{eqnarray}\label{I18}
\begin{split}
J_{H}&=-\frac{J_c^{2}}{4D\pi ^{2}R^{2}}\cos \left( \frac{J_0R_{y}}{D}%
\right) \sin \left( \frac{R\sqrt{J_0^{2}+4Du_{F}}}{D}\right), \\
J_{I}&=-\frac{J_c^{2}}{4D\pi ^{2}R^{2}}\left[ 1-\cos \left( \frac{J_0R_{y}%
}{D}\right) \right] \sin \left( \frac{R\sqrt{J_0^{2}+4Du_{F}}}{D}\right), \\
J_{DM}^{z}&=\frac{J_c^{2}}{4D\pi ^{2}R^{2}}\sin \left( \frac{J_0R_{y}}{D}%
\right) \sin \left( \frac{R\sqrt{J_0^{2}+4Du_{F}}}{D}\right).
\end{split}
\end{eqnarray}
As is evident from the above expressions, the RKKY interaction here comprise three contributions: the Heisenberg, Ising, and DM terms. Notably, both the orientation of the Ising term (along $z$) and the component of the DM term (the $z$-component) are governed by the N\'eel vector, in analogy to the $f$-wave case in the main text.
\par
To investigate the connection between the RKKY interaction and the odd-parity spin polarization, we fix the impurity distance $R$ and vary the azimuthal angle of the impurities to monitor the behavior of the RKKY components. As shown in Fig.~\ref{figS2}, the distance of each solid circle from the origin gives the magnitude of the RKKY component $J_i$; red (blue) and black indicate $J_i > 0$ ($<0$) and $J_i=0$, respectively. We find that, although all RKKY components exhibit a $p$-wave shape in real space, only the DM term displays odd parity, satisfying $J^{\alpha\beta}_{DM}(\mathbf{R}) =-J^{\alpha\beta}_{DM}(-C_{2q}\mathbf{R})$ ($q=1$), which directly characterizes the odd-parity spin polarization in momentum space, i.e., $S_z(\mathbf{k}) = -S_z(C_{2q}\mathbf{k})$.
\end{widetext}
%
\end{document}